\documentclass{pasj00}
\draft

\begin{document}
\SetRunningHead{Author(s) in page-head}{Running Head}

\title{The first four-color photometric investigation of the W UMa type contact binary V868 Mon}



%
 \author{Zhou \textsc{Xiao}\altaffilmark{1,2,3}
         Qian  \textsc{Shengbang}\altaffilmark{1,2,3}
         He \textsc{Jiajia}\altaffilmark{1,2}
         Zhang \textsc{Jia}\altaffilmark{1,2}
         Jiang \textsc{Linqiao}\altaffilmark{1,2,3}}
\altaffiltext{1}{Yunnan Observatories, Chinese Academy of Sciences, PO Box 110, 650216 Kunming, China}
 \email{zhouxiaophy@ynao.ac.cn}
\altaffiltext{2}{Key Laboratory of the Structure and Evolution of Celestial Objects, Chinese Academy of Sciences, PO Box 110, 650216 Kunming, China}
\altaffiltext{3}{Graduate University of the Chinese Academy of Sciences, Yuquan Road 19, Sijingshang Block, 100049 Beijing, China}


\maketitle

\begin{abstract}
The first four-color light curves of V868 Mon in the $B$ $V$ $R_c$ and $I_c$ bands are presented and analyzed by using the Wilson-Devinney method of the 2013 version. It is discovered that V868 Mon is an A-subtype contact binary (f=$58.9\,\%$) with a large temperature difference of 916$K$ between the two components. Using the eight new times of light minimum determined by the authors together with those collected from literatures, the authors found that the general trend of the observed-calculate ($O$-$C$) curve shows a upward parabolic variation that corresponds to a long-term increase in the orbital period at a rate of $dP/dt=9.38\times{10^{-7}}day\cdot year^{-1}$. The continuous increase may be caused by a mass transfer from the less massive component to the more massive one.
\end{abstract}

\section{Introduction}

V868 Mon, also named BD-2 2221 or GSC 4835 1947, is a relatively bright (V = 8.9 mag) eclipsing binary. Its light variability was first discovered on Stardial images by \citet{Wils2003}. The linear ephemeris determined by them is:
\begin{equation}
Min.I(HJD)=2452681.731+0^{d}.63772\times{E}.\label{linear ephemeris}
\end{equation}
Later, \citet{Otero2004} improved its orbital period to P = 0.637705 days and classified this binary as EB type. Then, \citet{Deb2011} listed the geometrical and physical parameters obtained from the $V$ band CCD data of the All Sky Automated Survey (ASAS)-3 project using the Wilson-Devinney code. As the ASAS-3 data of V868 Mon has a very low precision and was observed only in $V$ band, more observation about this target is still needed. The 2MASS infrared color index of V868 Mon gave $J$ - $K$ = 0.16, corresponding to an F0 spectral type. But the color index of $Tycho$-2 ($B$-$V$) = 0.20 redetermined the spectral type to be A8V. According to the spectral template result of \citet{Pribulla2009}, the spectral type between A8 to F1 was an acceptable result.

\section{New Photometric Observations}

\begin{figure}[!h]
\begin{center}
\includegraphics[width=16cm]{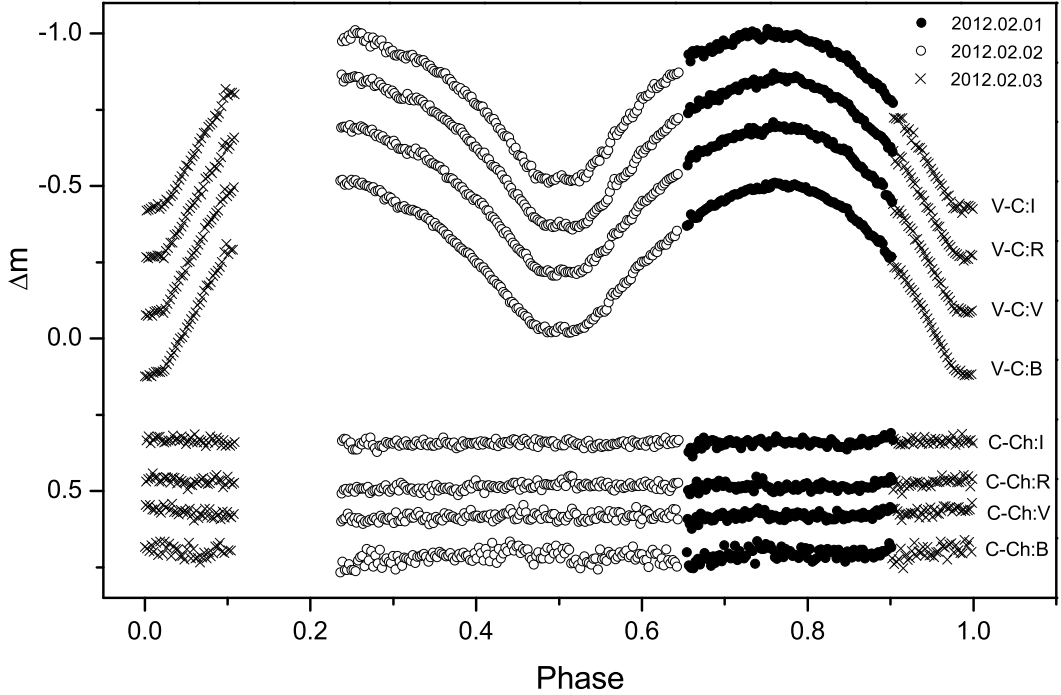}
\end{center}
\caption{CCD photometric light curves in the $B$ $V$ $R_c$ and $I_c$ bands obtained using the 60cm telescope at YNAO in 2012. The magnitude difference between the comparison and the check stars are also shown. Solid circles, open circles and crosses refer to the data observed on February 1, February 2 and February 3, respectively.} \label{fig1}
\end{figure}

The four-color ($B$ $V$ $R_c$ and $I_c$) light curves of V868 Mon were observed in three continuous nights on February 1, 2 and 3 in 2012 with the Andor DW436 2K CCD camera attached to the 60cm reflecting telescope at Yunnan Observatories (YNOs). The coordinates of the variable star, the comparison star and the check star are listed in Table \ref{Coordinates}. During the observation, the wide band, Johnson-Cousins $B$ $V$ $R_c$ $I_c$ filters were used. The integration time was 50s for $B$ band, 25s for $V$ band, 15s for $R_c$ band, and 10s for $I_c$ band respectively. A few lines of the original light curve data are displayed in Table \ref{data}. The phase of those observations displayed in Fig. \ref{fig1} were calculated with the following linear ephemeris:
\begin{equation}
Min.I(HJD)=2455961.142815125+0^{d}.637705\times{E}.\label{linear ephemeris}
\end{equation}

Meanwhile, the authors got three times of minima (TOMs) through the light curve observation on February 1, 2 and 3 in 2012.  After that, two TOMs were observed on February 28 and December 17 in 2012 using the 60cm reflecting telescope at Yunnan Observatories. One TOM was observed on January 21 in 2014 using the 85cm reflecting telescope in Xinglong Observation base. Two TOMs were observed on November 14 and 16 in 2014 using the 1m reflecting telescope at Yunnan Observatories. PHOT (measure magnitudes for a list of stars) of the aperture photometry package of the IRAF was used to reduce the observed images. By using a least-square parabolic fitting method, the new CCD times of light minimum were determined and listed in Table \ref{Newminimum}. The TOM observational data are list in Table 10 to Table 26.

\begin{table}[!h]
\begin{center}
\caption{Coordinates of V868 Mon, the comparison, and the check stars}\label{Coordinates}
\begin{small}
\begin{tabular}{cccc}\hline\hline
Targets          & name                      & $\alpha_{2000}$        & $\delta_{2000}$ \\\hline
Variable         & V868 Mon                  & $07^{h}39^{m}04.8^{s}$ & $-02^\circ39'05.6"$\\
The comparison   & 2MASS J07384996-0236215   & $07^{h}38^{m}49.9^{s}$ & $-02^\circ36'21.5"$\\
The check        & TYC 4835-2088-1           & $07^{h}38^{m}45.6^{s}$ & $-02^\circ37'52.5"$\\
\hline\hline
\end{tabular}
\end{small}
\end{center}
\end{table}

\begin{table}[!h]
\small
\caption{The original light curve data of V868 Mon observed by 60cm reflecting telescope in Yunnan Observatories}\label{data}
\begin{center}
\begin{tabular}{ccccccccccccc}\hline
JD(Hel.)    &              &     JD(Hel.)   &              &  JD(Hel.)      &              &    JD(Hel.)    &              &   JD(Hel.)     &              &   JD(Hel.)     &            \\
2455900+    &   $\Delta$m  &     2455900+   &   $\Delta$m  &  2455900+      & $\Delta$m    &     2455900+   & $\Delta$m    &   2455900+     &  $\Delta$m   &   2455900+     & $\Delta$m   \\\hline
59.00905	&    -0.369    &     59.02602	&    -0.430    &     59.04292	&    -0.464    &     59.05988	&    -0.487    &    59.07678	&    -0.511    &    59.09374	&    -0.501    \\
59.01078	&    -0.373    &     59.02768	&    -0.440    &     59.04464	&    -0.470    &     59.06154	&    -0.488    &    59.07850	&    -0.508    &    59.09540	&    -0.499    \\
59.01243	&    -0.407    &     59.02940	&    -0.436    &     59.04630	&    -0.473    &     59.06326	&    -0.495    &    59.08022	&    -0.509    &    59.09712	&    -0.496    \\
59.01416	&    -0.397    &     59.03106	&    -0.435    &     59.04802	&    -0.472    &     59.06498	&    -0.490    &    59.08188	&    -0.504    &    59.09884	&    -0.494    \\
59.01581	&    -0.395    &     59.03278	&    -0.442    &     59.04974	&    -0.475    &     59.06664	&    -0.491    &    59.08360	&    -0.500    &    59.10050	&    -0.496    \\
59.01754	&    -0.397    &     59.03443	&    -0.449    &     59.05140	&    -0.481    &     59.06836	&    -0.490    &    59.08526	&    -0.504    &    59.10222	&    -0.485    \\
59.01919	&    -0.410    &     59.03616	&    -0.453    &     59.05306	&    -0.481    &     59.07002	&    -0.506    &    59.08698	&    -0.495    &    59.10388	&    -0.482    \\
59.02092	&    -0.416    &     59.03781	&    -0.449    &     59.05478	&    -0.487    &     59.07174	&    -0.494    &    59.08864	&    -0.496    &    59.10560	&    -0.483    \\
59.02264	&    -0.421    &     59.03954	&    -0.455    &     59.05650	&    -0.489    &     59.07340	&    -0.496    &    59.09036	&    -0.503    &    59.10726	&    -0.482    \\
59.02430	&    -0.426    &     59.04126	&    -0.463    &     59.05816	&    -0.492    &     59.07512	&    -0.509    &    59.09202	&    -0.506    &    59.10898	&    -0.477    \\
\hline
\end{tabular}
\end{center}
\textbf
{\footnotesize Notes.} \footnotesize These are only a few lines of the light curve data, the whole data are list in Table 6, Table 7, Table 8 and Table 9.
\end{table}

\begin{table*}[!h]
\caption{$(O-C)$ values of light minimum times for V868 Mon}\label{Newminimum}
\begin{center}
\small
\begin{tabular}{llcllll}\hline\hline
JD (Hel.)       &   Error  &  Min     &   Epoch      & $(O-C)$         & Method    &  Reference       \\
(2400000+)      &          &          &              &                 &           &\\\hline
52681.731       &          & II       & -589.5       & 0.00590         &  CCD      &\citet{Wils2003}\\
53057.664       &          & I        & 0            & 0               &  CCD      &\citet{Otero2004}\\
53745.73825     &0.00063   & I        & 1079         & -0.00944        &  CCD      &\citet{Deb2011}\\
54157.0662      &0.00180   & I        & 1724         & -0.00122        &  CCD      &\citet{Otero2004}\\
55939.4617      &0.0002    & I        & 4519         & 0.00881         &  CCD      &\citet{Liakos2014}\\
55940.4192      &0.0002    & II       & 4520.5       & 0.00975         &  CCD      &\citet{Liakos2014}\\
55953.4923      &0.0002    & I        & 4541         & 0.00990         &  CCD      &\citet{Liakos2014}\\
55959.23184     &0.00021   & I        & 4550         & 0.01009         &  CCD      &60cm\\
55960.18838     &0.00029   & II       & 4551.5       & 0.01007         &  CCD      &60cm\\
55961.14282     &0.00020   & I        & 4553         & 0.00795         &  CCD      &60cm\\
55986.01873     &0.00096   & I        & 4592         & 0.01337         &  CCD      &60cm\\
55986.65190     &0.00090   & I        & 4593         & 0.00884         &  CCD      &\citet{Diethelm 2012}\\
56279.36496     &0.00079   & I        & 5052         & 0.01530         &  CCD      &60cm\\
56679.20920     &0.00023   & I        & 5679         & 0.01850         &  CCD      &85cm\\
56976.38424     &0.00017   & I        & 6145         & 0.02302         &  CCD      &1m\\
56978.29658     &0.00044   & I        & 6148         & 0.02224         &  CCD      &1m\\\hline
\end{tabular}
\end{center}
\textbf
{\footnotesize Notes.} \footnotesize 60cm and 1m denotes the 60cm and 1m reflecting telescope in
Yunnan Observatories respectively, and 85cm denotes to the 85cm reflecting telescope in Xinglong Observation base.
\end{table*}

\section{Orbital Period Variations}

The study of the orbital period change is a very important part for contact binary stars. But the period change investigation of V868 Mon has been neglected since it was discovered as an contact binary system in 2003. In this paper, we collected all available times of light minimum and list them in Table \ref{Newminimum}. Using the ephemeris given in O-C gateway \footnote{http://astro.sci.muni.cz/variables/ocgate/},
\begin{equation}
Min.I(HJD)=2453057.664+0^{d}.637705\times{E}.\label{linear ephemeris}
\end{equation}
the $(O - C)$ values are computed. The $(O-C)$ values (observational times of light minimum-calculational times of light minimum) calculated by equation
(3) are listed in the fourth column of Table 3, and plotted in the upper panel of Fig. 2. During the computation, times with the same epoch have been averaged, and only the mean values are listed in Table 3.
The general (O-C)$_{1}$ trend of V868 Mon shown in the upper panel of Fig. 3 indicates a visual change in its orbit. Based on the least-square method, an upward parabolic variation is added to the linear ephemeris of Equation (3). The new ephemeris is

\begin{equation}
\begin{array}{lll}
Min. I=2453057.65933(\pm0.00018)+0.637704468(\pm0.000000135)\times{E}
         \\+0.81956(\pm0.023212)\times{10^{-9}}\times{E^{2}}.
\end{array}
\end{equation}
With the quadratic term included in this ephemeris, a continuous period increase, at a rate of
$dP/dt=9.38\times{10^{-7}}day\cdot year^{-1}$
is determined. The residuals from equation (4) are showed in the lower panel of Fig. 2.

\begin{figure}[!h]
\begin{center}
\includegraphics[width=13cm]{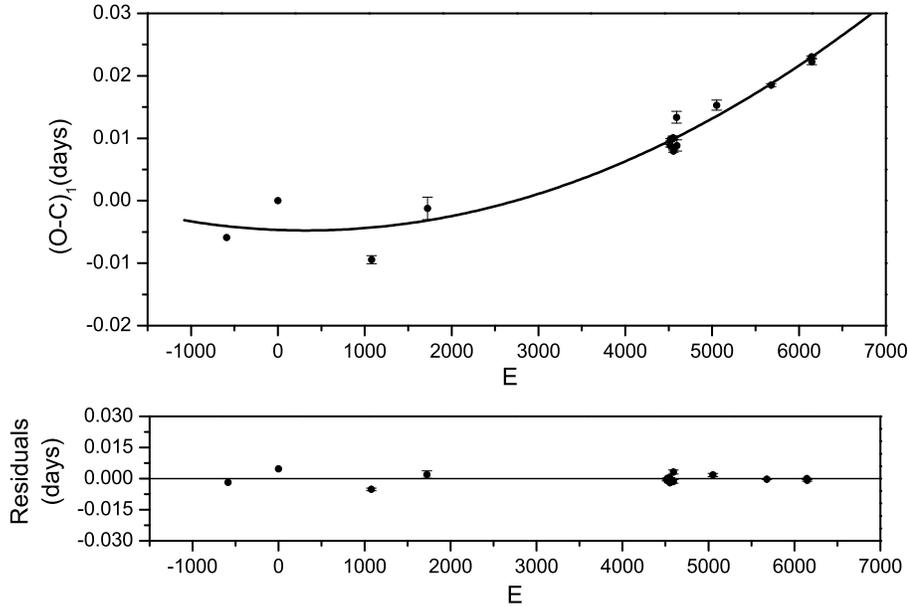}
\end{center}
\caption{A plot of the (O-C)$_{1}$ curve of V868 Mon form the linear ephemeris of Equation (3) is shown in the upper panel. The solid line in the panel refers to upward parabolic variation, which reveals a continuous increase in the orbital period. After upward parabolic change is removed, the residuals are plotted in the lower panel.}
\end{figure}

\section{The Analysis of the light curves by W-D Code}

V868 Mon has been found as a variable system for more than 10 years. To understand its geometrical structure and evolutionary state, the $B$ $V$ $R_c$ and $I_c$ light curves shown in Fig. 2 were analyzed using the W-D method of 2013 version (Wilson \& Devinney 1971; Wilson 1990, 1994; Wilson \& Van Hamme, 2003; Van Hamme \& Wilson 2007; Wilson 2008; Wilson et al. 2010; Wilson 2012). During the solution process, the effective temperature of star 1 was chosen as $T_1=7400$K according to its spectral type of A8/F1 (Cox2000). The radial velocity study of V868 Mon given by \citet{Pribulla2009} show that the mass ratio of this binary system was $q=0.373(8)$. As shown in Fig. 1, the light curve depths between the primary minima and the second minima have large difference, thus, the effective temperature between the primary star and the second star may also have large difference. So we take the values of the gravity-darkening coefficients and the bolometric albedo for both components differently, i.e., $g_{1}=1,g_{2}=0.32$ \citep{Lucy1967} and $A_{1}=1,A_{2}=0.5$ \citep{Rucinski1969}. The limb-darkening coefficients were used according to \citet{1993AJ....106.2096V}'s table (x and y are the bolometric and bandpass limb-darkening coefficients). The adjustable parameters were: the orbital inclination $i$; the mean temperature of star 2, $T_{2}$; the monochromatic luminosity of star 1, $L_{1B}$, $L_{1V}$, $L_{1R}$ and $L_{1I}$; the dimensionless potential of star 1 (mode 3 for overcontact configuration, $\Omega_{1}=\Omega_{2}$ ); and the third light, $l_{3}$. The photometric solutions are listed in Table \ref{phsolutions} and the theoretical light curves computed with those photometric parameters are plotted in Fig. 3 and Fig. 4. And also, the contact configuration of V868 Mon is displayed in Fig. 5.

\begin{table}[!h]
\caption{Photometric solutions for V868 Mon}\label{phsolutions}
\begin{center}
\small
\begin{tabular}{lllllllll}
\hline
Parameters                        & Without $L_{3}$               & With $L_{3}$           \\
\hline
$g_{1}$                           & 1.00(fixed)                   & 1.00(fixed)             \\
$g_{2}$                           & 0.32(fixed)                   & 0.32(fixed)             \\
$A_{1}$                           & 1.00(fixed)                   & 1.00(fixed)             \\
$A_{2}$                           & 0.50(fixed)                   & 0.50(fixed)             \\
q ($M_2/M_1$ )                    & 0.373(fixed)                  & 0.373(fixed)       \\
$T_{1}(K)   $                     & 7400(fixed)                   & 7400(fixed)             \\
$i(^{\circ})$                     & 78.52($\pm0.98$)              & 83.67($\pm0.16$)         \\
$\Omega_{in}$                     & 2.622506                      & 2.622506                 \\
$\Omega_{out}$                    & 2.393652                      & 2.393652                \\
$\Omega_{1}=\Omega_{2}$           & 2.487810                      & 2.517320                \\
$T_{2}(K)$                        & 6446($\pm7$)                  & 6484($\pm6$)             \\
$L_{1}/(L_{1}+L_{2}$) (B)         & 0.8344($\pm0.0008$)           & 0.8271($\pm0.0017$)                 \\
$L_{1}/(L_{1}+L_{2}$) (V)         & 0.8080($\pm0.0007$)           & 0.8011($\pm0.0022$)     \\
$L_{1}/(L_{1}+L_{2}$) (R)         & 0.7900($\pm0.0008$)           & 0.7832($\pm0.0033$)       \\
$L_{1}/(L_{1}+L_{2}$) (I)         & 0.7728($\pm0.0011$)           & 0.7662($\pm0.0052$)       \\
$r_{1}(pole)$                     & 0.4591($\pm0.0004$)           & 0.4652($\pm0.0006$)        \\
$r_{1}(side)$                     & 0.4965($\pm0.0006$)           & 0.5051($\pm0.0008$)        \\
$r_{1}(back)$                     & 0.5314($\pm0.0008$)           & 0.5432($\pm0.0012$)        \\
$r_{2}(pole)$                     & 0.2994($\pm0.0005$)           & 0.3062($\pm0.0007$)         \\
$r_{2}(side)$                     & 0.3156($\pm0.0006$)           & 0.3241($\pm0.0008$)         \\
$r_{2}(back)$                     & 0.3690($\pm0.0012$)           & 0.3872($\pm0.0019$)          \\
$L_{3}/(L_{1}+L_{2}+L_{3}$) (B)   &                               & 0.0263($\pm0.0001$)          \\
$L_{3}/(L_{1}+L_{2}+L_{3}$) (V)   &                               & 0.0224($\pm0.0001$)          \\
$L_{3}/(L_{1}+L_{2}+L_{3}$) (R)   &                               & 0.0193($\pm0.0002$)          \\
$L_{3}/(L_{1}+L_{2}+L_{3}$) (I)   &                               & 0.0175($\pm0.0001$)          \\
$f$                               & $46.0\,\%$($\pm$0.9\,\%$$)    &$58.9\,\%$($\pm$1.2\,\%$$)   \\
$\Sigma{\omega(O-C)^2}$           & 0.0163319                     &0.0113492                     \\
\hline
\end{tabular}
\end{center}
\textbf
{\footnotesize Notes.} \footnotesize The authors use the mass ratio of V868 Mon given by \citet{Pribulla2009} and made it as an unadjustable parameter.
\end{table}

\begin{figure}
\begin{center}
\includegraphics[width=14cm]{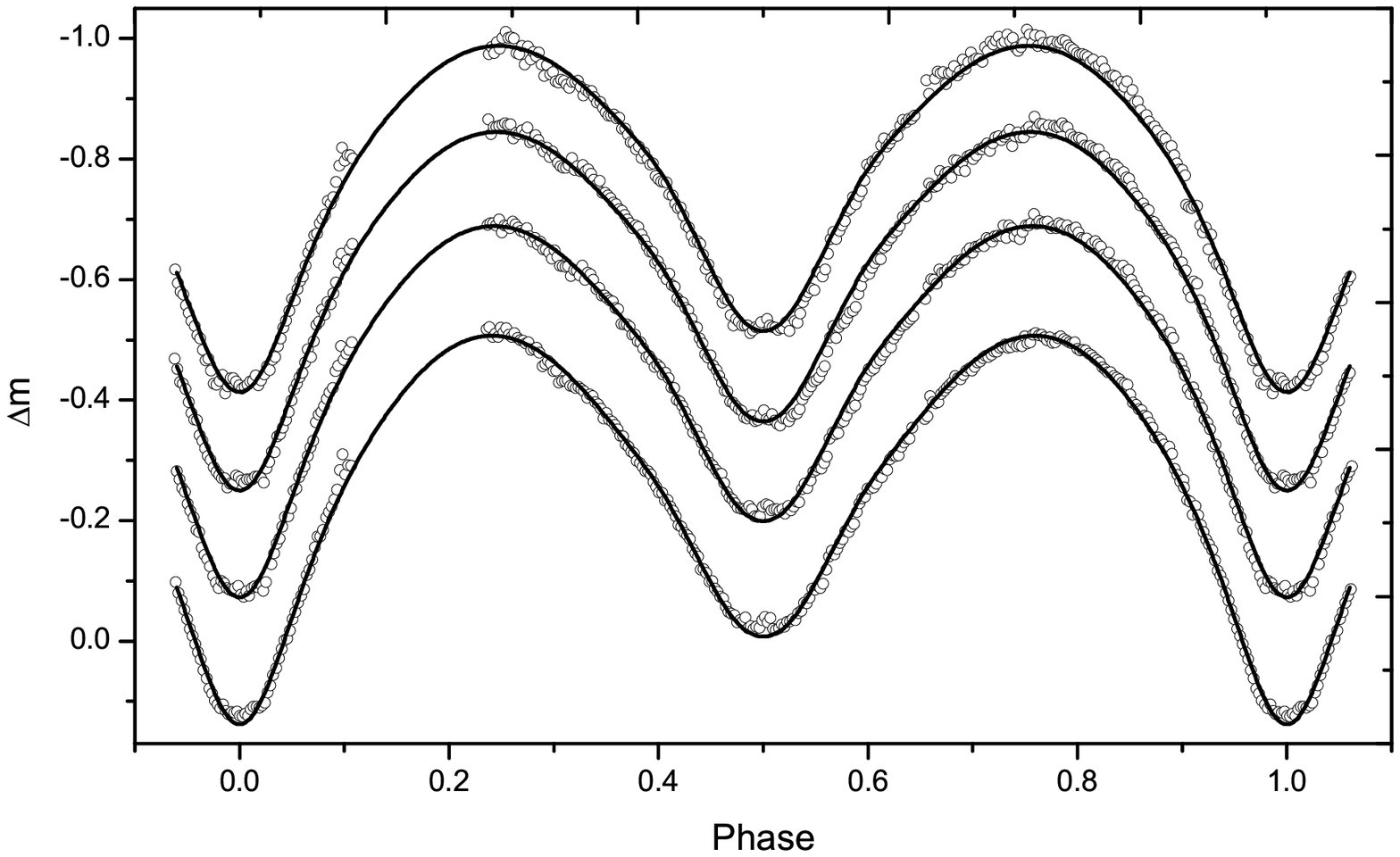}
\end{center}
\caption{Observed (open circles) and theoretical (line) light curves in the $B V R_c$ and $I_c$ band for V868 Mon (without $l_3$).}
\end{figure}

\begin{figure}
\begin{center}
\includegraphics[width=14cm]{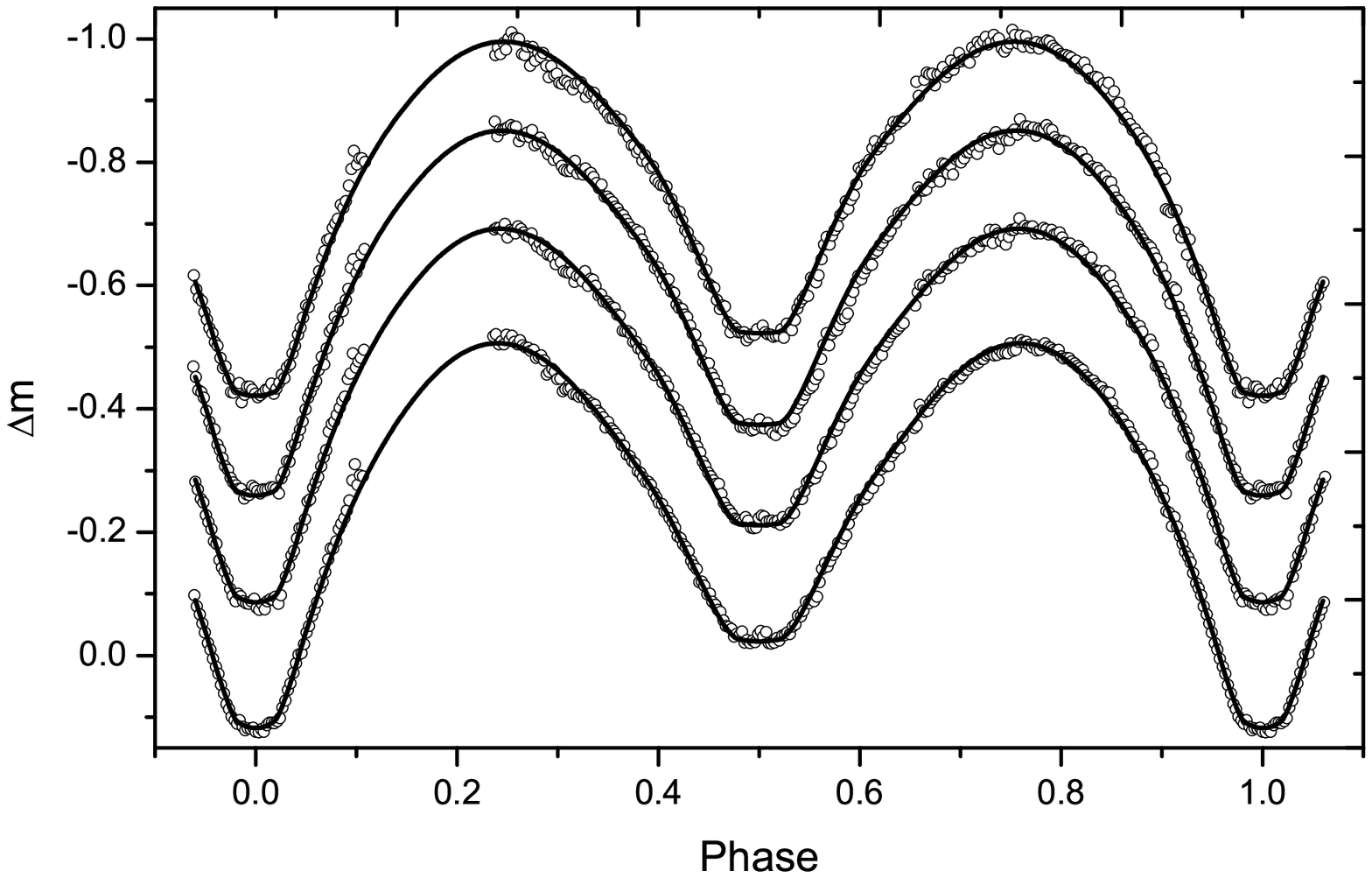}
\end{center}
\caption{Observed (open circles) and theoretical (line) light curves in the $B V R_c$ and $I_c$ band for V868 Mon (with $l_3$).}
\end{figure}

\begin{figure}
\begin{center}
\includegraphics[width=14cm]{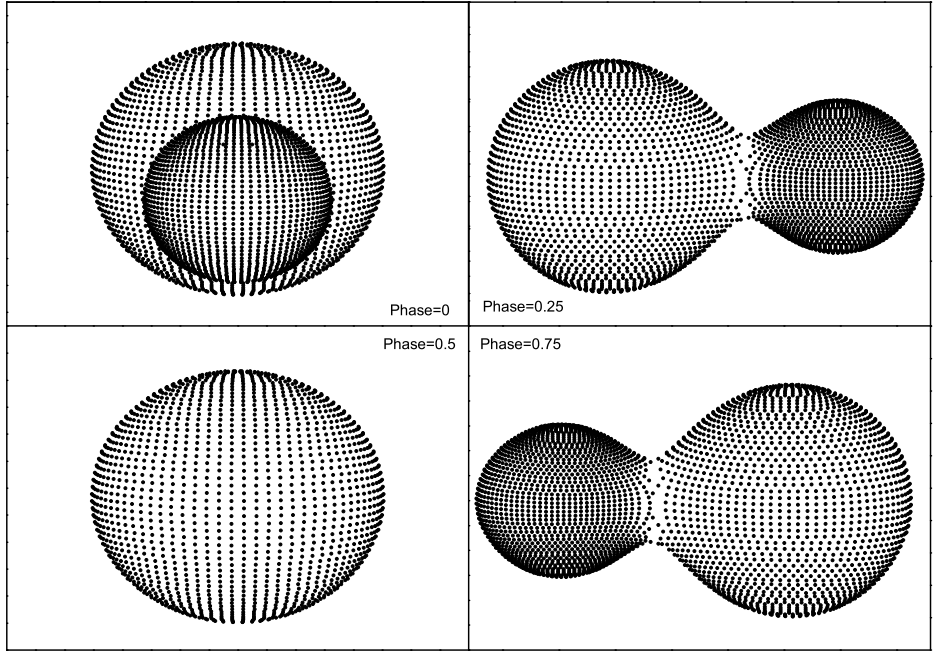}
\end{center}
\caption{Contact configurations of V868 Mon at phase 0.0, 0.25, 0.5, 0.75,}
\end{figure}

\section{Conclusions}

The light curve solution indicates that V868 Mon is a A-subtype
overcontact binary system. During the computing, the authors set the third light ($l_{3}$) as an adjustable parameter, and both of the results with and without $l_{3}$ are presented in Table \ref{phsolutions}. The results show that when $l_{3}$ is added, the W-D program will obtain a much smaller residual. In Fig. 3 and Fig. 4, the solid lines represent the theoretical light curves. It is obvious to point out that theoretical light curves containing third light give a much better fitting, especially for the primary and the second minima. Thus, V868 Mon is probably a triple system. The inclination of the binary system is $i=83.67^{\circ}$.
According to the radial velocity study by \citet{Pribulla2009}, who gave the result that $(M_1+M_2)sin^{3}i=2.504$
and q=0.373, the authors can get the absolute elements of V868 Mon in solar units as shown in Table \ref{elements}. The evolutionary statu of the primary star places it in the middle between the Zero Age Main Sequence (ZAMS) and Terminal Age Main Sequence (TAMS) lines of the H-R diagram. The secondary component is evidently more evolved than the primary star and is clearly overluminous and have higher effective temperature for its present mass.
\begin{table}[!h]
\caption{Absolute elements of  V868 Mon}\label{elements}
\begin{center}
\small
\begin{tabular}{lllllllll}
\hline
Parameters                           &Primary                 & Secondary          \\
\hline
$M$                               & $2.389(\pm0.509)M_\odot$       & $0.891(\pm0.190)M_\odot$         \\
$R$                               & $2.331(\pm0.001)R_\odot$       & $1.562(\pm0.001)R_\odot$         \\
$L$                               & $14.702(\pm0.010)L_\odot$      & $3.893(\pm0.005)L_\odot$         \\
\hline
\end{tabular}
\end{center}
\end{table}

The orbital period of V868 Mon is increasing at a rate of $dP/dt=9.38\times{10^{-7}}day\cdot year^{-1}$. The long term period decreasing
may be due to conservative mass transfer from the less massive component to the more massive one, which coincides with the
result of \citet{Qian2001}. By combining the absolute parameters of V868 Mon
with the well-known equation  $\frac{dM_{2}}{dt}=\frac{M_1M_2}{3p(M_1-M_2)}\times{dp/dt}$, the mass transfer, at a rate of
$\frac{dM_{2}}{dt}=6.93\times{10^{-7}}M_\odot/year$ is determined. However, we can't exclude the possibility of angular momentum loss due to a magnetic stellar wind. As V868 Mon is a late A, early F type star, and our observation have not detected any activity of magnetic spot, so maybe the
mass transfer from the less massive component to the more massive one is more reliable. And also, as the light curves analysis shows that there is a third component around the binary system and the O-C diagram consists of only 16 points in a time span of 12 years, the upward parabolic variation may be just a part of the periodic variation caused by the third component. Thus, more observation of TOMs are needed in the future.

Most of contact binaries have temperatures of the components roughly equal because of sharing a common envelope with the same entropy, thereby making the effective temperatures almost equal over the surface \citep{Paczy2006}. As in the case of V868 Mon, the contact of degree is $58.9\,\%$, which means this system is a deep contact binary system, and photometric solutions list in Table 4 gives a large temperature difference (916K) between the primary and the secondary star. The authors suppose that V868 Mon is a detach system at first, then mass transfer occurs between the two components, and it evolve into a contact system gradually. This is quite different from those of late type contact binaries in which the existence of the third component have played an important role during their formation and evolution \citep{Qian2013}. The formation and evolution of early-type contact binaries are still unsolved problems in stellar astrophysics. More observation and analysis about early type contact binaries are quite needed.

In summary, V868 Mon is a deep contact (contact of degree larger than $50\,\%$) binary with total eclipses, which allows us to derive very precise photometric elements. Long term period investigation reveals that the period of V868 Mon shows a secular period increase, which may due to the mass transfer between the two components.

\begin{table}[!h]
\small
\begin{center}
\caption{The original data of V868 Mon in $B$ band observed by 60cm}
\begin{tabular}{ccccccccccccc}\hline
JD(Hel.)   &               &     JD(Hel.)      &             &  JD(Hel.)     &           &    JD(Hel.)    &            &   JD(Hel.)      &             &   JD(Hel.)     &            \\
2455900+   &   $\Delta$m   &     2455900+      &   $\Delta$m &  2455900+     & $\Delta$m &     2455900+   & $\Delta$m  &   2455900+      &  $\Delta$m  &   2455900+     & $\Delta$m   \\\hline
59.00905   &	   -0.369  &    59.10222	   &   -0.485    &    60.04494	 &   -0.483  &     60.13811	  &    -0.180  &    60.23128	 &    -0.153   &    61.76368    &	0.087     \\
59.01078   &	   -0.373  &    59.10388	   &   -0.482    &    60.04666	 &   -0.485  &     60.13977	  &    -0.176  &    60.23300     &    -0.163   &    61.76534    &	0.099     \\
59.01243   &	   -0.407  &    59.10560       &   -0.483    &    60.04832	 &   -0.487  &     60.14149	  &    -0.169  &    60.23466	 &    -0.169   &    61.76706    &	0.105     \\
59.01416   &	   -0.397  &    59.10726	   &   -0.482    &    60.05004	 &   -0.479  &     60.14321	  &    -0.153  &    60.23638	 &    -0.183   &    61.76879    &	0.111     \\
59.01581   &	   -0.395  &    59.10898	   &   -0.477    &    60.05177	 &   -0.456  &     60.14487	  &    -0.146  &    60.23810     &    -0.181   &    61.77044    &	0.105     \\
59.01754   &	   -0.397  &    59.11064	   &   -0.474    &    60.05342	 &   -0.460  &     60.14659	  &    -0.140  &    60.23976	 &    -0.190   &    61.77217    &	0.116     \\
\hline
\end{tabular}
\end{center}
\end{table}

\addtocounter{table}{-1}
\begin{table}[!h]
\small
\begin{center}
\caption{Continuted}
\begin{tabular}{ccccccccccccccc}\hline
JD(Hel.)   &               &     JD(Hel.)    &             &  JD(Hel.)     &           &    JD(Hel.)    &            &   JD(Hel.)    &             &  JD(Hel.)      &            \\
2455900+   &   $\Delta$m   &     2455900+     &$\Delta$m   &  2455900+      &  $\Delta$m  &      2455900+   &   $\Delta$m   &   2455900+     &   $\Delta$m  &   2455900+      &  $\Delta$m \\\hline
59.07340   &	   -0.496  &    59.16657	   &   -0.268    &    60.10929	 &   -0.299  &     60.20246	  &    -0.033  &    61.73492	 &    -0.151   &    61.19032    &	-0.174    \\
59.07512   &	   -0.509  &    60.01790	   &   -0.518    &    60.11101	 &   -0.296  &     60.20418	  &    -0.037  &    61.73658	 &    -0.134   &    61.19205    &	-0.174    \\
59.07678   &	   -0.511  &    60.01956	   &   -0.521    &    60.11273	 &   -0.289  &     60.20584	  &    -0.035  &    61.73830	 &    -0.117   &    61.19370    & 	-0.185  \\
59.07850   &	   -0.508  &    60.02128	   &   -0.505    &    60.11439	 &   -0.282  &     60.20756	  &    -0.047  &    61.73996	 &    -0.109   &    61.19543    &	-0.205    \\
59.08022   &	   -0.509  &    60.02294	   &   -0.510    &    60.11611	 &   -0.282  &     60.20928	  &    -0.050  &    61.74168	 &    -0.098   &    61.19708    &	-0.214    \\
59.08188   &	   -0.504  &    60.02466	   &   -0.505    &    60.11777	 &   -0.277  &     60.21094	  &    -0.064  &    61.74334	 &    -0.080   &    61.19881    &	-0.223    \\
\hline
\end{tabular}
\end{center}
\end{table}

\begin{table}[!h]
\small
\begin{center}
\caption{The original data of V868 Mon in $V$ band observed by 60cm}
\begin{tabular}{ccccccccccccc}\hline
JD(Hel.)   &               &     JD(Hel.)      &             &  JD(Hel.)       &           &    JD(Hel.)    &              &   JD(Hel.)    &             &  JD(Hel.)      &            \\
2455900+   &    $\Delta$m  &     2455900+      &  $\Delta$m  &  2455900+       & $\Delta$m &      2455900+  &  $\Delta$m   &   2455900+    &  $\Delta$m  &   2455900+ & $\Delta$m     \\\hline
59.00963   &    	-0.568 &     59.10280      &    -0.678   &  60.04552	   &  -0.663   &       60.13868	&       -0.380 &   60.23185    &  	-0.345   &      61.76426  &   	-0.108 \\
59.01135   &    	-0.589 &     59.10445	   &    -0.669   &  60.04724	   &  -0.669   &       60.14034	&       -0.368 &   60.23358    &  	-0.363   &      61.76592  &   	-0.097 \\
59.01301   &    	-0.601 &     59.10618	   &    -0.667   &  60.04890	   &  -0.658   &       60.14206	&       -0.359 &   60.23523    &  	-0.365   &      61.76764  &   	-0.088 \\
59.01473   &    	-0.600 &     59.10783	   &    -0.668   &  60.05062	   &  -0.663   &       60.14372	&       -0.346 &   60.23696    &  	-0.374   &      61.76930  &     -0.100   \\
59.01639   &    	-0.593 &     59.10956	   &    -0.655   &  60.05228	   &  -0.650   &       60.14544	&       -0.351 &   60.23861    &  	-0.388   &      61.77102  &   	-0.096 \\
59.01811   &    	-0.617 &     59.11121	   &    -0.660   &  60.05400	   &  -0.650   &       60.14717	&       -0.339 &   60.24034    &  	-0.387   &      61.77268  &   	-0.088 \\
\hline
\end{tabular}
\end{center}
\end{table}

\addtocounter{table}{-1}
\begin{table}[!h]
\small
\begin{center}
\caption{Continuted}
\begin{tabular}{ccccccccccccccc}\hline
JD(Hel.)   &               &     JD(Hel.)      &             &  JD(Hel.)       &           &      JD(Hel.)  &              &   JD(Hel.)    &             &  JD(Hel.)      &            \\
2455900+   &   $\Delta$m   &     2455900+      &$\Delta$m    &  2455900+       & $\Delta$m &      2455900+  &  $\Delta$m   &   2455900+    & $\Delta$m   &   2455900+     & $\Delta$m \\\hline
59.07397   &    	-0.694 &     59.16708	   &    -0.448   &  60.10986	   &  -0.498   &       60.20303	&       -0.217 &   61.73543    &  	-0.324   &      61.19090  &    -0.376   \\
59.07569   &    	-0.709 &     60.01841	   &    -0.691   &  60.11158	   &  -0.496   &       60.20475	&       -0.223 &   61.73716    &  	-0.326   &      61.19262  &   	-0.380  \\
59.07735   &    	-0.693 &     60.02014	   &    -0.694   &  60.11324	   &  -0.486   &       60.20641	&       -0.232 &   61.73888    &  	-0.306   &      61.19428  &   	-0.390  \\
59.07907   &    	-0.697 &     60.02186	   &    -0.694   &  60.11496	   &  -0.472   &       60.20813	&       -0.231 &   61.74054    &  	-0.296   &      61.19600  &     -0.402  \\
59.08073   &    	-0.694 &     60.02352	   &    -0.690   &  60.11668	   &  -0.473   &       60.20985	&       -0.252 &   61.74226    &  	-0.282   &      61.19766  &   	-0.413 \\
59.08245   &    	-0.695 &     60.02524	   &    -0.700   &  60.11834	   &  -0.466   &       60.21151	&       -0.261 &   61.74392    &  	-0.269   &      61.19938  &   	-0.420  \\
\hline
\end{tabular}
\end{center}
\end{table}

\begin{table}[!h]
\small
\begin{center}
\caption{The original data of V868 Mon in $R$ band observed by 60cm}
\begin{tabular}{ccccccccccccc}\hline
JD(Hel.)    &               &     JD(Hel.) &             &  JD(Hel.)    &           &    JD(Hel.)    &            &   JD(Hel.)    &             &  JD(Hel.)      &            \\
2455900+    & $\Delta$m     &     2455900+ & $\Delta$m   &  2455900+    &   $\Delta$m   &      2455900+   &   $\Delta$m   &   2455900+     &  $\Delta$m     &   2455900+      & $\Delta$m \\\hline
59.01001    &   	-0.739  &  59.10312    &   	-0.835  &   60.04590    &   	-0.831  &    60.13900    &   	-0.553  &   60.23224    &   	-0.521  &    61.76458    &   	-0.288  \\
59.01167    &   	-0.759  &  59.10484    &   	-0.833  &   60.04756    &   	-0.840  &    60.14073    &   	-0.549  &   60.23389    &   	-0.527  &    61.76630    &   	-0.281  \\
59.01339    &   	-0.754  &  59.10650    &   	-0.823  &   60.04928    &   	-0.832  &    60.14245    &   	-0.535  &   60.23562    &   	-0.528  &    61.76802    &   	-0.276  \\
59.01505    &   	-0.754  &  59.10822    &   	-0.821  &   60.05100    &   	-0.822  &    60.14411    &   	-0.522  &   60.23727    &   	-0.536  &    61.76968    &   	-0.270  \\
59.01677    &   	-0.756  &  59.10988    &   	-0.822  &   60.05266    &   	-0.813  &    60.14583    &   	-0.520  &   60.23900    &   	-0.546  &    61.77140    &   	-0.270  \\
59.01843    &   	-0.780  &  59.11160    &   	-0.818  &   60.05438    &   	-0.819  &    60.14749    &   	-0.502  &   60.24072    &   	-0.548  &    61.77306    &   	-0.255  \\
\hline
\end{tabular}
\end{center}
\end{table}

\addtocounter{table}{-1}
\begin{table}[!h]
\small
\begin{center}
\caption{Continuted}
\begin{tabular}{ccccccccccccccc}\hline
JD(Hel.)   &             &     JD(Hel.)    &             &  JD(Hel.)     &           &    JD(Hel.)    &            &   JD(Hel.)    &             &  JD(Hel.)      &            \\
2455900+    &$\Delta$m   &     2455900+     & $\Delta$m       &  2455900+      & $\Delta$m  &      2455900+   &   $\Delta$m   &   2455900+     &    $\Delta$m  &   2455900+      & $\Delta$m  \\\hline
59.07436    &   	-0.845  &  59.16746    &   	-0.612  &   60.11024    &   	-0.687  &    60.20341    &   	-0.373  &   61.73582    &   	-0.507  &    61.19128    &   	-0.530  \\
59.07601    &   	-0.870  &  60.01880    &   	-0.866  &   60.11196    &   	-0.675  &    60.20513    &   	-0.378  &   61.73754    &   	-0.482  &    61.19294    &   	-0.550  \\
59.07774    &   	-0.852  &  60.02052    &   	-0.842  &   60.11362    &   	-0.666  &    60.20679    &   	-0.378  &   61.73920    &   	-0.471  &    61.19466    &   	-0.542  \\
59.07939    &   	-0.859  &  60.02218    &   	-0.853  &   60.11534    &   	-0.655  &    60.20851    &   	-0.385  &   61.74092    &   	-0.469  &    61.19632    &   	-0.563  \\
59.08111    &   	-0.858  &  60.02390    &   	-0.853  &   60.11700    &   	-0.647  &    60.21017    &   	-0.401  &   61.74258    &   	-0.452  &    61.19804    &   	-0.559  \\
59.08277    &   	-0.851  &  60.02556    &   	-0.855  &   60.11872    &   	-0.646  &    60.21189    &   	-0.409  &   61.74430    &       -0.430  &    61.19976    &   	-0.577  \\
\hline
\end{tabular}
\end{center}
\end{table}

\begin{table}[!h]
\small
\begin{center}
\caption{The original data of V868 Mon in $I$ band observed by 60cm}
\begin{tabular}{ccccccccccccc}\hline
JD(Hel.)    &                &     JD(Hel.) &                &  JD(Hel.)        &                &    JD(Hel.)  &                &   JD(Hel.)   &                &  JD(Hel.)    &            \\
2455900+    &  $\Delta$m     &  2455900+    & $\Delta$m      &  2455900+        &  $\Delta$m     &    2455900+  &     $\Delta$m  &   2455900+   &    $\Delta$m   &   2455900+   & $\Delta$m \\\hline
59.01027	&     -0.931     &   59.10337	&     -0.973     &       60.04615	&     -0.978     &    60.13932	&     -0.690     &    60.23421	&     -0.690     &   61.76655	&     -0.426     \\
59.01192	&     -0.908     &   59.10509	&     -0.974     &       60.04788	&     -0.971     &    60.14098	&     -0.686     &    60.23587	&     -0.697     &   61.76828	&     -0.432     \\
59.01365	&     -0.943     &   59.10681	&     -0.972     &       60.04953	&     -0.977     &    60.14270	&     -0.673     &    60.23759	&     -0.714     &   61.76993	&     -0.442     \\
59.01537	&     -0.934     &   59.10847	&     -0.970     &       60.05126	&     -0.957     &    60.14436	&     -0.663     &    60.23925	&     -0.718     &   61.77166	&     -0.411     \\
59.01703	&     -0.945     &   59.11019	&     -0.962     &       60.05298	&     -0.938     &    60.14608	&     -0.664     &    60.24097	&     -0.737     &   61.77338	&     -0.423     \\
59.01875	&     -0.943     &   59.11185	&     -0.965     &       60.05464	&     -0.956     &    60.14780	&     -0.659     &    60.24263	&     -0.725     &   61.77504	&     -0.436     \\
\hline
\end{tabular}
\end{center}
\end{table}

\addtocounter{table}{-1}
\begin{table}[!h]
\small
\begin{center}
\caption{Continuted}
\begin{tabular}{ccccccccccccccc}\hline
JD(Hel.)    &                &     JD(Hel.) &                &  JD(Hel.)        &                &    JD(Hel.)  &                &   JD(Hel.)   &                &  JD(Hel.)    &            \\
2455900+    & $\Delta$m      &     2455900+ & $\Delta$m      &  2455900+        &  $\Delta$m     &    2455900+  &  $\Delta$m     &   2455900+   &     $\Delta$m  &   2455900+   &  $\Delta$m \\\hline
59.07461	&     -0.993     &   59.16772	&     -0.773     &       60.11050	&     -0.821     &    60.20539	&     -0.531     &    61.73779	&     -0.626     &   61.19319	&     -0.701     \\
59.07633	&     -1.006     &   60.01905	&     -0.974     &       60.11222	&     -0.802     &    60.20705	&     -0.528     &    61.73945	&     -0.626     &   61.19492	&     -0.708     \\
59.07799	&     -0.988     &   60.02077	&     -0.986     &       60.11388	&     -0.811     &    60.20877	&     -0.544     &    61.74117	&     -0.617     &   61.19664	&     -0.731     \\
59.07971	&     -1.004     &   60.02243	&     -0.976     &       60.11560	&     -0.800     &    60.21049	&     -0.563     &    61.74290	&     -0.594     &   61.19830	&     -0.726     \\
59.08137	&     -0.993     &   60.02415	&     -0.994     &       60.11726	&     -0.791     &    60.21215	&     -0.573     &    61.74455	&     -0.580     &   61.20002	&     -0.737     \\
59.08309	&     -0.986     &   60.02587	&     -0.983     &       60.11898	&     -0.793     &    60.21387	&     -0.579     &    61.74628	&     -0.575     &   61.20168	&     -0.762     \\
\hline
\end{tabular}
\end{center}
\end{table}

\begin{table}[!h]
\small
\caption{The TOM observational data of V868 Mon in $B$ band observed by 60cm}
\begin{center}
\begin{tabular}{ccccccccccccc}\hline
JD(Hel.)     &           &     JD(Hel.)    &             &  JD(Hel.)        &            &    JD(Hel.)     &             &    JD(Hel.)     &  \\
2455900+     & $\Delta$m &  2455900+       & $\Delta$m   &  2455900+        &  $\Delta$m &    2455900+     &  $\Delta$m  &    2455900+     &     $\Delta$m\\\hline
85.997139    &	-1.083   &    86.051340    &	-1.225   &     86.103829    &	-1.540   &    86.156329    &	-1.665   &    86.211445    &	-1.616   \\
85.998840    &	-1.087   &    86.053030    &	-1.230   &     86.105519    &	-1.552   &    86.158019    &	-1.655   &    86.213146    &	-1.597   \\
86.000530    &	-1.059   &    86.054720    &	-1.259   &     86.107220    &	-1.564   &    86.159720    &	-1.682   &    86.214847    &	-1.621   \\
86.002220    &	-1.063   &    86.056410    &	-1.272   &     86.108910    &	-1.551   &    86.161410    &	-1.708   &    86.216537    &	-1.589   \\
86.003921    &	-1.039   &    86.058111    &	-1.300   &     86.110600    &	-1.564   &    86.165704    &	-1.670   &    86.218239    &	-1.568   \\
86.005611    &	-1.053   &    86.059801    &	-1.294   &     86.112301    &	-1.552   &    86.167394    &	-1.652   &    86.219928    &	-1.566   \\
\hline
\end{tabular}
\end{center}
\end{table}

\begin{table}[!h]
\small
\caption{The TOM observational data of V868 Mon in $V$ band observed by 60cm}
\begin{center}
\begin{tabular}{ccccccccccccc}\hline
JD(Hel.)     &             &     JD(Hel.)  &             &  JD(Hel.)         &             &    JD(Hel.)       &             &    JD(Hel.)     &              \\
2455900+     & $\Delta$m   &  2455900+     & $\Delta$m   &  2455900+         &  $\Delta$m  &    2455900+       &  $\Delta$m  &    2455900+     &     $\Delta$m\\\hline
85.997712	   &     -0.357  &  86.051902	   &     -0.498  &      86.104390	   &     -0.801  &     86.156890	   &     -0.890  &    86.212017	   &     -0.878  \\
85.999402	   &     -0.365  &  86.053592	   &     -0.509  &      86.106080	   &     -0.802  &     86.158580	   &     -0.890  &    86.213707	   &     -0.855  \\
86.001092	   &     -0.357  &  86.055281	   &     -0.520  &      86.107781	   &     -0.805  &     86.160281	   &     -0.929  &    86.215409	   &     -0.882  \\
86.002781	   &     -0.332  &  86.056983	   &     -0.536  &      86.109471	   &     -0.823  &     86.161971	   &     -0.969  &    86.217110	   &     -0.861  \\
86.004483	   &     -0.347  &  86.058673	   &     -0.556  &      86.111173	   &     -0.811  &     86.166265	   &     -0.915  &    86.218800	   &     -0.847  \\
86.006173	   &     -0.337  &  86.060362	   &     -0.554  &      86.112862	   &     -0.816  &     86.167955	   &     -0.906  &    86.220490	   &     -0.828  \\
\end{tabular}
\end{center}
\end{table}

\begin{table}[!h]
\small
\caption{The TOM observational data of V868 Mon in $R$ band observed by 60cm}
\begin{center}
\begin{tabular}{ccccccccccccc}\hline
JD(Hel.)     &           &     JD(Hel.)    &             &  JD(Hel.)        &            &    JD(Hel.)     &             &    JD(Hel.)     &  \\
2455900+     & $\Delta$m &  2455900+       & $\Delta$m   &  2455900+        &  $\Delta$m &    2455900+     &  $\Delta$m  &    2455900+     &     $\Delta$m\\\hline
85.998071     &  	0.185     &    86.052261     &  	0.076     &   86.106450     &  	-0.219    &   86.158950     &  	-0.335    &   86.214078     &  	-0.280    \\
85.999761     &  	0.213     &    86.053950     &  	0.063     &   86.108140     &  	-0.229    &   86.160640     &  	-0.383    &   86.215767     &  	-0.310    \\
86.001450     &  	0.205     &    86.055640     &  	0.052     &   86.109830     &  	-0.232    &   86.162330     &  	-0.440    &   86.217469     &  	-0.292    \\
86.003152     &  	0.229     &    86.057342     &  	0.032     &   86.111531     &  	-0.226    &   86.166624     &  	-0.331    &   86.219159     &  	-0.259    \\
86.004842     &  	0.214     &    86.059031     &  	0.026     &   86.113221     &  	-0.228    &   86.168314     &  	-0.327    &   86.220860     &  	-0.278    \\
86.006531     &  	0.198     &    86.060721     &  	0.027     &   86.114911     &  	-0.252    &   86.170004     &  	-0.329    &   86.222550     &  	-0.280    \\
\hline
\end{tabular}
\end{center}
\end{table}

\begin{table}[!h]
\small
\caption{The TOM observational data of V868 Mon in $I$ band observed by 60cm}
\begin{center}
\begin{tabular}{ccccccccccccc}\hline
JD(Hel.)     &           &     JD(Hel.)    &             &  JD(Hel.)        &            &    JD(Hel.)     &             &    JD(Hel.)     &  \\
2455900+     & $\Delta$m &  2455900+       & $\Delta$m   &  2455900+        &  $\Delta$m &    2455900+     &  $\Delta$m  &    2455900+     &     $\Delta$m\\\hline
85.998343   &   	0.690     &     86.052533   &   	0.574     &     86.106722   &   	0.292     &    86.159222   &   	0.132     &   86.214350   &   	0.212     \\
86.000033   &   	0.713     &     86.054222   &   	0.563     &     86.108412   &   	0.287     &    86.160912   &   	0.124     &   86.216039   &   	0.202     \\
86.001734   &   	0.715     &     86.055924   &   	0.538     &     86.110114   &   	0.287     &    86.162614   &   	0.127     &   86.217741   &   	0.232     \\
86.003424   &   	0.706     &     86.057614   &   	0.533     &     86.111803   &   	0.280     &    86.166896   &   	0.184     &   86.219431   &   	0.272     \\
86.005114   &   	0.713     &     86.059303   &   	0.540     &     86.113493   &   	0.268     &    86.168586   &   	0.215     &   86.221132   &   	0.259     \\
86.006815   &   	0.725     &     86.060993   &   	0.513     &     86.115183   &   	0.259     &    86.170287   &   	0.192     &   86.222822   &   	0.249     \\
\hline
\end{tabular}
\end{center}
\end{table}

\begin{table}[!h]
\small
\caption{The TOM observational data of V868 Mon in $I$ band observed by 60cm}
\begin{center}
\begin{tabular}{ccccccccccccc}\hline
JD(Hel.)     &           &     JD(Hel.)    &             &  JD(Hel.)        &            &    JD(Hel.)     &             \\
2456200+     & $\Delta$m &  2456200+       & $\Delta$m   &  2456200+        &  $\Delta$m &    2456200+     &  $\Delta$m  \\\hline
79.308954	  &      0.437 &     79.329359	  &      0.600 &    79.349764	  &      0.698 &     79.368804	  &      0.751 \\
79.310320	  &      0.438 &     79.330725	  &      0.610 &    79.351130	  &      0.698 &     79.370169	  &      0.735 \\
79.311686	  &      0.462 &     79.332079	  &      0.630 &    79.352484	  &      0.710 &     79.371535	  &      0.737 \\
79.313040	  &      0.469 &     79.333445	  &      0.643 &    79.353850	  &      0.710 &     79.372889	  &      0.735 \\
79.314406	  &      0.483 &     79.334811	  &      0.661 &    79.355204	  &      0.716 &     79.374244	  &      0.742 \\
79.315760	  &      0.493 &     79.336165	  &      0.658 &    79.356570	  &      0.727 &     79.375609	  &      0.734 \\
\hline
\end{tabular}
\end{center}
\end{table}

\begin{table}[!h]
\small
\caption{The TOM observational data of V868 Mon in $N$ band observed by 60cm }
\begin{center}
\begin{tabular}{ccccccccccccc}\hline
JD(Hel.)     &           &     JD(Hel.)    &             &  JD(Hel.)        &            &    JD(Hel.)     &             \\
2456200+     & $\Delta$m &  2456200+       & $\Delta$m   &  2456200+        &  $\Delta$m &    2456200+     &  $\Delta$m  \\\hline
79.309122       &     -0.003  &    79.329527       &     0.175   &    79.349932       &     0.279   &    79.368972       &     0.307   \\
79.310488       &     -0.01   &    79.330893       &     0.173   &    79.351298       &     0.264   &    79.370337       &     0.330   \\
79.311842       &     0.013   &    79.332247       &     0.179   &    79.352652       &     0.283   &    79.371691       &     0.310   \\
79.313208       &     0.016   &    79.333613       &     0.202   &    79.354006       &     0.287   &    79.373057       &     0.332   \\
79.314562       &     0.026   &    79.334967       &     0.195   &    79.355372       &     0.293   &    79.374411       &     0.297   \\
79.315928       &     0.043   &    79.336333       &     0.198   &    79.356726       &     0.295   &    79.375777       &     0.304   \\
\hline
\end{tabular}
\end{center}
\end{table}

\begin{table}[!h]
\small
\caption{The TOM observational data of V868 Mon in $V$ band observed by 85cm}
\begin{center}
\begin{tabular}{ccccccccccccc}\hline
JD(Hel.)     &           &     JD(Hel.)    &             &  JD(Hel.)        &            &    JD(Hel.)     &             &    JD(Hel.)    \\
2456600+     & $\Delta$m &  2456600+       & $\Delta$m   &  2456600+        &  $\Delta$m &    2456600+     &  $\Delta$m  &    2456600+       $\Delta$m\\\hline
    79.13596 & -0.674 & 79.15727 & -0.677 & 79.17671 & -0.527 & 79.19724 & -0.383 & 79.21669 & -0.371 \\
    79.13751 & -0.786 & 79.15789 & -0.668 & 79.17734 & -0.516 & 79.19787 & -0.379 & 79.21732 & -0.379 \\
    79.13820 & -0.778 & 79.15853 & -0.670 & 79.17796 & -0.517 & 79.19850 & -0.379 & 79.21794 & -0.383 \\
    79.13927 & -0.775 & 79.15916 & -0.661 & 79.17860 & -0.521 & 79.19912 & -0.377 & 79.21857 & -0.389 \\
    79.13989 & -0.766 & 79.15978 & -0.662 & 79.17923 & -0.508 & 79.19975 & -0.380 & 79.21919 & -0.389 \\
    79.14034 & -0.772 & 79.16041 & -0.653 & 79.17985 & -0.502 & 79.20037 & -0.372 & 79.21983 & -0.387 \\
\hline
\end{tabular}
\end{center}
\end{table}

\begin{table}[!h]
\small
\caption{The TOM observational data of V868 Mon in $R$ band observed by 85cm}
\begin{center}
\begin{tabular}{ccccccccccccc}\hline
JD(Hel.)     &           &     JD(Hel.)    &             &  JD(Hel.)        &            &    JD(Hel.)     &             &    JD(Hel.)    \\
2456600+     & $\Delta$m &  2456600+       & $\Delta$m   &  2456600+        &  $\Delta$m &    2456600+     &  $\Delta$m  &    2456600+       $\Delta$m\\\hline
    79.13635 & -0.166 & 79.15810 & -0.077 & 79.17755 & 0.076 & 79.19808 & 0.208 & 79.21752 & 0.219 \\
    79.13842 & -0.178 & 79.15873 & -0.060 & 79.17817 & 0.086 & 79.19871 & 0.203 & 79.21815 & 0.213 \\
    79.13948 & -0.174 & 79.15936 & -0.069 & 79.17881 & 0.093 & 79.19933 & 0.209 & 79.21877 & 0.209 \\
    79.14055 & -0.166 & 79.15999 & -0.060 & 79.17943 & 0.096 & 79.19996 & 0.201 & 79.21940 & 0.210 \\
    79.14117 & -0.170 & 79.16061 & -0.049 & 79.18006 & 0.101 & 79.20058 & 0.205 & 79.22004 & 0.202 \\
    79.14180 & -0.157 & 79.16124 & -0.053 & 79.18068 & 0.102 & 79.20122 & 0.218 & 79.22066 & 0.204 \\
\hline
\end{tabular}
\end{center}
\end{table}

\begin{table}[!h]
\small
\caption{The TOM observational data of V868 Mon in $I$ band observed by 85cm}
\begin{center}
\begin{tabular}{ccccccccccccc}\hline
JD(Hel.)     &           &     JD(Hel.)    &             &  JD(Hel.)        &            &    JD(Hel.)     &             &    JD(Hel.)     \\
2456600+     & $\Delta$m &  2456600+       & $\Delta$m   &  2456600+        &  $\Delta$m &    2456600+     &  $\Delta$m  &    2456600+       $\Delta$m\\\hline
    79.13666 & 0.269 & 79.15830 & 0.453 & 79.17774 & 0.594 & 79.19828 & 0.716 & 79.21772 & 0.725 \\
    79.13862 & 0.348 & 79.15892 & 0.462 & 79.17837 & 0.602 & 79.19890 & 0.725 & 79.21835 & 0.721 \\
    79.13967 & 0.360 & 79.15956 & 0.458 & 79.17901 & 0.600 & 79.19953 & 0.710 & 79.21897 & 0.713 \\
    79.14074 & 0.350 & 79.16019 & 0.477 & 79.17963 & 0.613 & 79.20015 & 0.727 & 79.21960 & 0.722 \\
    79.14137 & 0.356 & 79.16081 & 0.475 & 79.18026 & 0.610 & 79.20078 & 0.719 & 79.22023 & 0.736 \\
    79.14199 & 0.368 & 79.16144 & 0.480 & 79.18088 & 0.630 & 79.20141 & 0.723 & 79.22086 & 0.719 \\
\hline
\end{tabular}
\end{center}
\end{table}

\begin{table}[!h]
\small
\caption{The TOM observational data of V868 Mon in $V$ band observed by 1m}
\begin{center}
\begin{tabular}{ccccccccccccc}\hline
JD(Hel.)     &           &     JD(Hel.)    &             &  JD(Hel.)        &            &    JD(Hel.)     &             &    JD(Hel.)     \\
2456900+     & $\Delta$m &  2456900+       & $\Delta$m   &  2456900+        &  $\Delta$m &    2456900+     &  $\Delta$m  &    2456900+        $\Delta$m\\\hline
    76.32173 & -0.702 & 76.35049 & -0.495 & 76.37684 & -0.340 & 76.40320 & -0.378 & 76.42955 & -0.590 \\
    76.32296 & -0.691 & 76.35136 & -0.496 & 76.37772 & -0.336 & 76.40408 & -0.383 & 76.43043 & -0.599 \\
    76.32419 & -0.685 & 76.35224 & -0.488 & 76.37860 & -0.341 & 76.40496 & -0.395 & 76.43131 & -0.603 \\
    76.32542 & -0.677 & 76.35312 & -0.477 & 76.37948 & -0.345 & 76.40583 & -0.396 & 76.43219 & -0.605 \\
    76.32665 & -0.672 & 76.35400 & -0.468 & 76.38036 & -0.339 & 76.40671 & -0.408 & 76.43307 & -0.618 \\
    76.32763 & -0.666 & 76.35488 & -0.458 & 76.38123 & -0.344 & 76.40759 & -0.410 & 76.43395 & -0.621 \\
\hline
\end{tabular}
\end{center}
\end{table}

\begin{table}[!h]
\small
\caption{The TOM observational data of V868 Mon in $R$ band observed by 1m}
\begin{center}
\begin{tabular}{ccccccccccccc}\hline
JD(Hel.)     &           &     JD(Hel.)    &             &  JD(Hel.)        &            &    JD(Hel.)     &             &    JD(Hel.)     \\
2456900+     & $\Delta$m &  2456900+       & $\Delta$m   &  2456900+        &  $\Delta$m &    2456900+     &  $\Delta$m  &    2456900+        $\Delta$m\\\hline
    76.32200 & -0.128 & 76.35076 & 0.067 & 76.37799 & 0.208 & 76.40435 & 0.170 & 76.43071 & -0.037 \\
    76.32323 & -0.126 & 76.35164 & 0.074 & 76.37887 & 0.216 & 76.40523 & 0.157 & 76.43158 & -0.041 \\
    76.32446 & -0.113 & 76.35252 & 0.075 & 76.37975 & 0.215 & 76.40611 & 0.159 & 76.43246 & -0.042 \\
    76.32569 & -0.107 & 76.35340 & 0.090 & 76.38063 & 0.217 & 76.40699 & 0.151 & 76.43334 & -0.050 \\
    76.32692 & -0.103 & 76.35428 & 0.093 & 76.38151 & 0.215 & 76.40786 & 0.144 & 76.43422 & -0.059 \\
    76.32791 & -0.096 & 76.35515 & 0.106 & 76.38239 & 0.207 & 76.40874 & 0.137 & 76.43510 & -0.065 \\
\hline
\end{tabular}
\end{center}
\end{table}

\begin{table}[!h]
\small
\caption{The TOM observational data of V868 Mon in $I$ band observed by 1m}
\begin{center}
\begin{tabular}{ccccccccccccc}\hline
JD(Hel.)     &           &     JD(Hel.)    &             &  JD(Hel.)        &            &    JD(Hel.)     &             &    JD(Hel.)    \\
2456900+     & $\Delta$m &  2456900+       & $\Delta$m   &  2456900+        &  $\Delta$m &    2456900+     &  $\Delta$m  &    2456900+       $\Delta$m\\\hline
    76.32221 & 0.399 & 76.35098 & 0.594 & 76.37821 & 0.724 & 76.40457 & 0.685 & 76.43092 & 0.490 \\
    76.32345 & 0.406 & 76.35185 & 0.598 & 76.37909 & 0.729 & 76.40545 & 0.684 & 76.43180 & 0.488 \\
    76.32467 & 0.408 & 76.35273 & 0.608 & 76.37997 & 0.726 & 76.40632 & 0.674 & 76.43268 & 0.478 \\
    76.32591 & 0.419 & 76.35361 & 0.612 & 76.38084 & 0.732 & 76.40720 & 0.668 & 76.43356 & 0.474 \\
    76.32714 & 0.428 & 76.35449 & 0.620 & 76.38172 & 0.729 & 76.40808 & 0.660 & 76.43443 & 0.465 \\
    76.32812 & 0.433 & 76.35537 & 0.624 & 76.38260 & 0.726 & 76.40896 & 0.663 & 76.43532 & 0.457 \\
\hline
\end{tabular}
\end{center}
\end{table}

\begin{table}[!h]
\small
\caption{The TOM observational data of V868 Mon in $N$ band observed by 1m}
\begin{center}
\begin{tabular}{ccccccccccccc}\hline
JD(Hel.)     &           &     JD(Hel.)    &             &  JD(Hel.)        &            &    JD(Hel.)     &             &    JD(Hel.)     \\
2456900+     & $\Delta$m &  2456900+       & $\Delta$m   &  2456900+        &  $\Delta$m &    2456900+     &  $\Delta$m  &    2456900+       $\Delta$m\\\hline
    76.32237 & -0.096 & 76.35114 & 0.103 & 76.37837 & 0.240 & 76.40473 & 0.200 & 76.43108 & -0.004 \\
    76.32361 & -0.088 & 76.35202 & 0.102 & 76.37925 & 0.238 & 76.40561 & 0.185 & 76.43196 & -0.013 \\
    76.32484 & -0.086 & 76.35290 & 0.113 & 76.38013 & 0.243 & 76.40648 & 0.190 & 76.43284 & -0.019 \\
    76.32607 & -0.085 & 76.35377 & 0.122 & 76.38101 & 0.235 & 76.40736 & 0.175 & 76.43372 & -0.028 \\
    76.32730 & -0.071 & 76.35465 & 0.128 & 76.38189 & 0.240 & 76.40824 & 0.169 & 76.43460 & -0.028 \\
    76.32828 & -0.059 & 76.35553 & 0.136 & 76.38277 & 0.240 & 76.40912 & 0.158 & 76.43548 & -0.042 \\
\hline
\end{tabular}
\end{center}
\end{table}

\begin{table}[!h]
\small
\caption{The TOM observational data of V868 Mon in $V$ band observed by 1m}
\begin{center}
\begin{tabular}{ccccccccccccc}\hline
JD(Hel.)     &           &     JD(Hel.)    &             &  JD(Hel.)        &            &    JD(Hel.)     &            \\
2456900+     & $\Delta$m &  2456900+       & $\Delta$m   &  2456900+        &  $\Delta$m &    2456900+     &  $\Delta$m     $\Delta$m\\\hline
    78.25663 & -0.552 & 78.28028 & -0.358 & 78.29995 & -0.327 & 78.32022 & -0.403 \\
    78.25864 & -0.528 & 78.28197 & -0.348 & 78.30164 & -0.328 & 78.32191 & -0.417 \\
    78.26062 & -0.518 & 78.28366 & -0.339 & 78.30333 & -0.331 & 78.32360 & -0.430 \\
    78.26231 & -0.496 & 78.28535 & -0.337 & 78.30502 & -0.320 & 78.32529 & -0.417 \\
    78.26400 & -0.481 & 78.28704 & -0.332 & 78.30671 & -0.333 & 78.32698 & -0.460 \\
    78.26569 & -0.470 & 78.28872 & -0.327 & 78.30840 & -0.337 & 78.32866 & -0.467 \\
    78.26738 & -0.460 & 78.29041 & -0.333 & 78.31009 & -0.332 & 78.33186 & -0.498 \\
    78.26906 & -0.443 & 78.29319 & -0.327 & 78.31178 & -0.347 & 78.33355 & -0.517 \\
    78.27244 & -0.415 & 78.29488 & -0.325 & 78.31347 & -0.352 & 78.33524 & -0.522 \\
    78.27413 & -0.398 & 78.29657 & -0.329 & 78.31684 & -0.372 & 78.33693 & -0.532 \\
    78.27582 & -0.381 & 78.29826 & -0.328 & 78.31853 & -0.398 & 78.33862 & -0.554 \\
    78.27859 & -0.364 &       &       &       &       &       &  \\
\hline
\end{tabular}
\end{center}
\end{table}

\begin{table}[!h]
\small
\caption{The TOM observational data of V868 Mon in $R$ band observed by 1m}
\begin{center}
\begin{tabular}{ccccccccccccc}\hline
JD(Hel.)     &           &     JD(Hel.)    &             &  JD(Hel.)        &            &    JD(Hel.)     &            \\
2456900+     & $\Delta$m &  2456900+       & $\Delta$m   &  2456900+        &  $\Delta$m &    2456900+     &  $\Delta$m     $\Delta$m\\\hline
    78.25722 & 0.020 & 78.28087 & 0.202 & 78.30055 & 0.224 & 78.32081 & 0.150 \\
    78.25935 & 0.038 & 78.28256 & 0.214 & 78.30224 & 0.221 & 78.32250 & 0.140 \\
    78.26121 & 0.042 & 78.28425 & 0.219 & 78.30393 & 0.219 & 78.32419 & 0.121 \\
    78.26290 & 0.067 & 78.28594 & 0.221 & 78.30562 & 0.220 & 78.32757 & 0.091 \\
    78.26459 & 0.080 & 78.28763 & 0.211 & 78.30730 & 0.217 & 78.33246 & 0.062 \\
    78.26628 & 0.097 & 78.28932 & 0.224 & 78.30899 & 0.212 & 78.33415 & 0.042 \\
    78.27134 & 0.135 & 78.29379 & 0.221 & 78.31068 & 0.216 & 78.33584 & 0.031 \\
    78.27303 & 0.135 & 78.29548 & 0.223 & 78.31237 & 0.204 & 78.33752 & 0.020 \\
    78.27472 & 0.168 & 78.29717 & 0.219 & 78.31406 & 0.177 &       &  \\
    78.27918 & 0.193 & 78.29886 & 0.225 & 78.31912 & 0.157 &       &  \\
\hline
\end{tabular}
\end{center}
\end{table}

\begin{table}[!h]
\small
\caption{The TOM observational data of V868 Mon in $I$ band observed by 1m}
\begin{center}
\begin{tabular}{ccccccccccccc}\hline
JD(Hel.)     &           &     JD(Hel.)    &             &  JD(Hel.)        &            &    JD(Hel.)     &            \\
2456900+     & $\Delta$m &  2456900+       & $\Delta$m   &  2456900+        &  $\Delta$m &    2456900+     &  $\Delta$m     $\Delta$m\\\hline
    78.25758 & 0.545 & 78.27954 & 0.710 & 78.29922 & 0.732 & 78.31948 & 0.676 \\
    78.25977 & 0.568 & 78.28123 & 0.712 & 78.30091 & 0.719 & 78.32117 & 0.660 \\
    78.26158 & 0.581 & 78.28292 & 0.726 & 78.30259 & 0.727 & 78.32286 & 0.641 \\
    78.26326 & 0.591 & 78.28461 & 0.727 & 78.30428 & 0.732 & 78.32624 & 0.629 \\
    78.26495 & 0.602 & 78.28630 & 0.721 & 78.30598 & 0.734 & 78.32792 & 0.620 \\
    78.26664 & 0.615 & 78.28799 & 0.728 & 78.30766 & 0.722 & 78.33282 & 0.572 \\
    78.26833 & 0.665 & 78.28968 & 0.730 & 78.30935 & 0.723 & 78.33451 & 0.552 \\
    78.27170 & 0.658 & 78.29415 & 0.730 & 78.31104 & 0.724 & 78.33620 & 0.551 \\
    78.27339 & 0.665 & 78.29584 & 0.728 & 78.31442 & 0.707 & 78.33788 & 0.533 \\
    78.27508 & 0.679 & 78.29753 & 0.722 & 78.31610 & 0.702 &       &  \\
\hline
\end{tabular}
\end{center}
\end{table}

\begin{table}[!h]
\small
\caption{The TOM observational data of V868 Mon in $N$ band observed by 1m}
\begin{center}
\begin{tabular}{ccccccccccccc}\hline
JD(Hel.)     &           &     JD(Hel.)    &             &  JD(Hel.)        &            &    JD(Hel.)     &            \\
2456900+     & $\Delta$m &  2456900+       & $\Delta$m   &  2456900+        &  $\Delta$m &    2456900+     &  $\Delta$m     $\Delta$m\\\hline
    78.25782 & 0.097 & 78.28148 & 0.274 & 78.30115 & 0.292 & 78.31973 & 0.221 \\
    78.26002 & 0.117 & 78.28317 & 0.292 & 78.30285 & 0.284 & 78.32142 & 0.201 \\
    78.26182 & 0.130 & 78.28486 & 0.280 & 78.30454 & 0.279 & 78.32311 & 0.182 \\
    78.26351 & 0.156 & 78.28655 & 0.275 & 78.30622 & 0.274 & 78.32480 & 0.177 \\
    78.26520 & 0.136 & 78.28824 & 0.290 & 78.30791 & 0.268 & 78.32649 & 0.156 \\
    78.26689 & 0.168 & 78.28993 & 0.283 & 78.30960 & 0.269 & 78.32818 & 0.136 \\
    78.26858 & 0.153 & 78.29440 & 0.272 & 78.31129 & 0.265 & 78.33307 & 0.101 \\
    78.27195 & 0.208 & 78.29609 & 0.280 & 78.31298 & 0.249 & 78.33476 & 0.094 \\
    78.27364 & 0.227 & 78.29778 & 0.283 & 78.31467 & 0.257 & 78.33645 & 0.096 \\
    78.27533 & 0.240 & 78.29946 & 0.3   & 78.31636 & 0.243 & 78.33813 & 0.071 \\
    78.27979 & 0.267 &       &       &       &       &       &  \\
\hline
\end{tabular}
\end{center}
\end{table}

\bigskip

\vskip 0.3in \noindent
This work is supported by the Chinese Natural Science Foundation (Grant No. 11133007 and 11325315), the Strategic Priority Research Program ``The Emergence of Cosmological Structure'' of the Chinese Academy of Sciences (Grant No. XDB09010202) and the Science Foundation of Yunnan Province (Grant No. 2012HC011). New CCD photometric observations of V868 Mon were obtained with the 60cm and the 1.0m telescopes at the Yunnan Observatories, and the 85cm telescope in Xinglong Observation base in China.



\begin{thebibliography}{}
\bibitem[Cox(2000)]{Cox2000} Cox, A. N. 2000, Allen¡¯s Astrophysical Quantities (4th ed.; NewYork: Springer)
\bibitem[Deb et al.(2011)]{Deb2011} Deb, S., \& Singh, H.~P.\ 2011, \mnras, 412, 1787
\bibitem[Diethelm et al.(2012)]{Diethelm 2012} Diethelm, R.\ 2012,IBVS, 6029, 1
\bibitem[Liakos et al.(2014)]{Liakos2014} Liakos, A., Gazeas, K.,\& Nanouris, N.\ 2014, IBVS, 6095, 1
\bibitem[Lucy(1967)]{Lucy1967} Lucy, L. B., 1967, Zert. Astrophys. 65, 89
\bibitem[Otero et al.(2004)]{Otero2004} Otero, S.~A., Wils, P.,\& Dubovsky, P.~A.\ 2004, IBVS, 5570, 1
\bibitem[Paczy{\'n}ski et al.(2006)]{Paczy2006} Paczy{\'n}ski,B., Szczygie{\l}, D.~M., Pilecki, B.,\& Pojma{\'n}ski, G.\ 2006, \mnras, 368, 1311
\bibitem[Pribulla et al.(2009)]{Pribulla2009} Pribulla, T.,Rucinski, S.~M., Blake, R.~M., et al.\ 2009, \aj, 137, 3655
\bibitem[Qian et al.(2001)]{Qian2001} Qian, S.-B.,\ 2001, \mnras, 328, 635
\bibitem[Qian et al.(2013)]{Qian2013} Qian, S.-B., Liu, N.-P.,Li, K., et al.\ 2013, \apjs, 209, 13
\bibitem[Rucinski et al.(1969)]{Rucinski1969} Rucinski, S. M., 1969, A\&A 19, 245
\bibitem[Van Hamme(1993)]{1993AJ....106.2096V} Van Hamme, W.\ 1993, \aj, 106, 2096
\bibitem[Van et al.(2007)]{Van2007} Van Hamme, W., \& Wilson, R.~E.\ 2007, \apj, 661, 1129
\bibitem[Wils et al.(2003)]{Wils2003} Wils, P., \& Dvorak, S.~W.\ 2003, IBVS, 5425, 1
\bibitem[Wilson(1990)]{Wilson1990} Wilson, R. E. 1990, ApJ, 356, 613
\bibitem[Wilson(1994)]{Wilson1994} Wilson, R. E., 1994, PASP 106, 921
\bibitem[Wilson(2003)]{Wilson2003} Wilson, R.~E.\ 2008, \apj, 672,575
\bibitem[Wilson(2012)]{Wilson2012} Wilson, R.~E.\ 2012, \aj, 144,73
\bibitem[Wilson et al.(1971)]{Wilson1971} Wilson, R. E. \& Devinney, E. J., 1971, ApJ 166, 605
\bibitem[Wilson(2003)]{Wilson2003} Wilson, R. E. \& Van Hamme, W., 2003, Computing Binary Stars Observables, the 4th edition of the W-D program
\bibitem[Wilson et al(2010)]{Wilson2010} Wilson, R.~E., Van Hamme, W., \& Terrell, D.\ 2010, \apj, 723, 1469
\end{thebibliography}
\end{document}